\documentclass[10pt,letterpaper]{article}
\usepackage{opex3}

\usepackage[latin1]{inputenc}
\usepackage[english]{babel}
\usepackage{cite}
\usepackage{graphicx}
\usepackage{amsmath}
\usepackage{amssymb}
\usepackage{units}
\usepackage{color}
\usepackage{float}
\usepackage[caption=false]{subfig} 

\DeclareMathOperator{\Var}{Var}
\newcommand{\op}[1]{\hat{#1}}
\newcommand{\fiftyfifty}{$50$\,:\,$50$ }

\begin{document}

\title{Stable control of 10 dB two-mode squeezed vacuum states of light}

\author{Tobias Eberle,$^1$ Vitus H\"andchen,$^1$ and Roman Schnabel$^{1,*}$}
\address{$^1$Max-Planck-Institut f\"ur Gravitationsphysik
(Albert-Einstein-Institut) and\\ Institut f\"ur Gravitationsphysik
der Leibniz Universit\"at Hannover, Callinstraße 38, 30167 Hannover,
Germany}
\email{*roman.schnabel@aei.mpg.de}

\begin{abstract}
    Continuous variable entanglement is a fundamental resource for many quantum information tasks. Important protocols like superactivation of zero-capacity channels and finite-size quantum cryptography that provides security against most general attacks, require about \unit[10]{dB} two-mode squeezing. Additionally, stable phase control mechanisms are necessary but are difficult to achieve because the total amount of optical loss to the entangled beams needs to be small. Here, we experimentally demonstrate a control scheme for two-mode squeezed vacuum states at the telecommunication wavelength of \unit[1550]{nm}. Our states exhibited an Einstein-Podolsky-Rosen covariance product of $0.0309 \pm 0.0002$, where $1$ is the critical value, and a Duan inseparability value of $0.360 \pm 0.001$, where $4$ is the critical value. The latter corresponds to $\unit[10.45 \pm 0.01]{dB}$ which reflects the average non-classical noise suppression of the two squeezed vacuum states used to generate the entanglement. With the results of this work demanding quantum information protocols will become feasible.
\end{abstract}

\ocis{(270.6570) Squeezed states; (270.5565) Quantum communications; (270.5568) Quantum cryptography.}

\section{Introduction}

Since the foundation of quantum mechanics entanglement has been proven to be a valuable resource in quantum information tasks and has spread a variety of applications~\cite{Horodecki2009} ranging from teleportation~\cite{Bouwmeester1997,Furusawa1998} and quantum dense coding~\cite{Bennett1992,Braunstein2000} to quantum dense metrology~\cite{Schnabel2010,Wasilewski2010,Steinlechner2012} and quantum cryptography~\cite{Weedbrook2012}. It is also an important ingredient to quantum repeaters~\cite{Briegel1998} and quantum computation~\cite{DiVincenzo1995}.

Quantum cryptography has already matured and first commercial systems exist using observables with either a discrete or a continuous eigenspectrum. The security of commercially available continuous variable (CV) systems relies on security proofs under the assumption of collective attacks~\cite{Lodewyck2007}. Only recently a security proof for most general coherent attacks with a finite number of samples providing composable security was published~\cite{Furrer2012}. The protocol uses two-mode squeezed vacuum states and homodyne measurements. It reaches positive key rates only for high degree of two-mode squeezing, low channel loss and a moderately large number of samples in the order of $10^8$. To achieve these requirements a stable control of the entanglement generation is necessary.

A proposal for continuous variable superactivation~\cite{Smith2008} of two zero-capacity channels was recently published~\cite{Smith2011}. Superactivation describes the effect that when two channels with a capacity of zero are combined, e.g.\ a $\unit[50]{\%}$ loss channel and a channel with positive partial transpose, the resulting capacity will become positive. The proposed scheme to reach a positive channel capacity involves amongst others two-mode squeezed vacuum states with stably controlled $\unit[10]{dB}$ entanglement.

\begin{figure}
  \center
  \includegraphics[width=8.5cm]{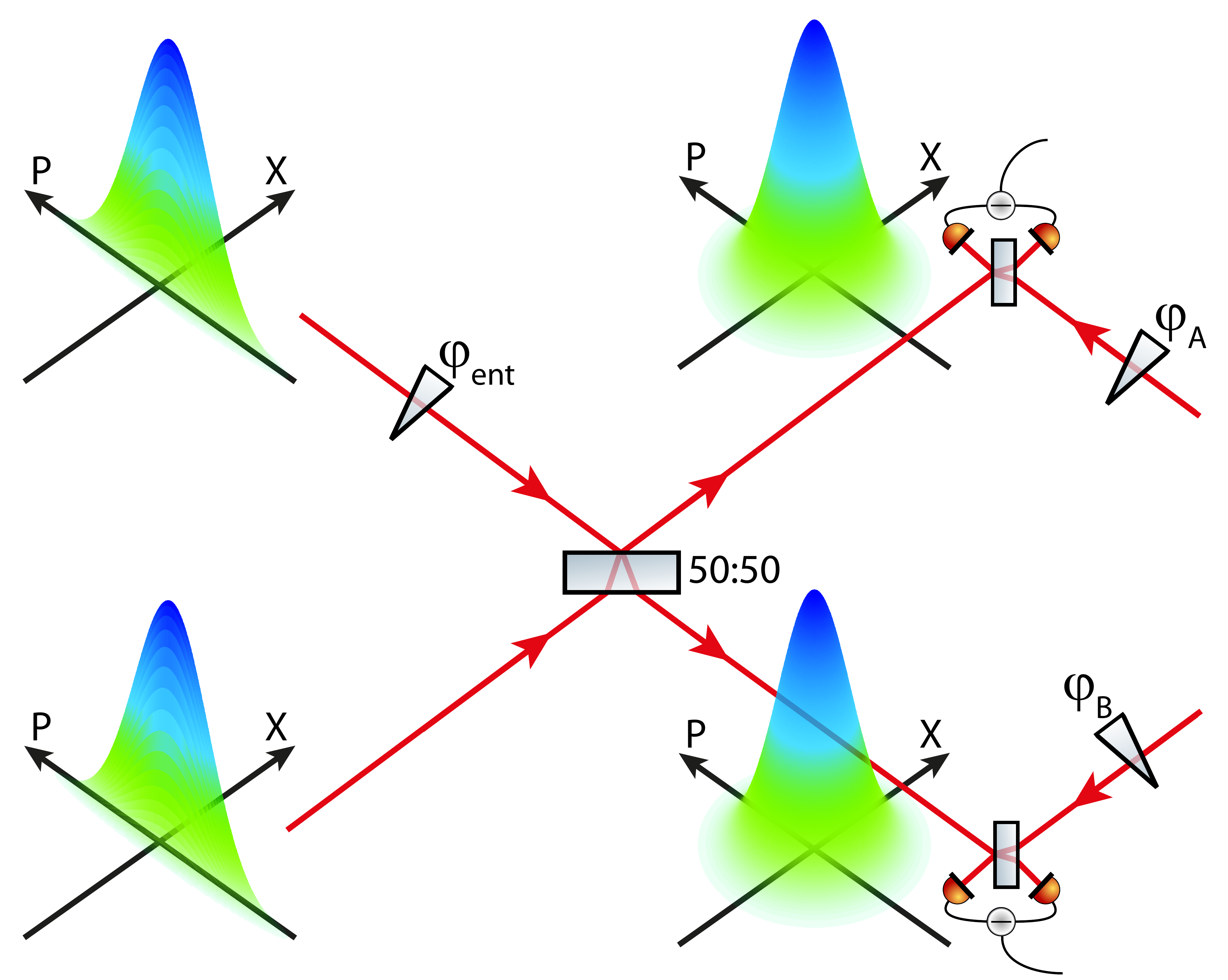}
  \caption{Principle of generating and detecting two-mode squeezing. Two squeezed vacuum modes are superimposed at a $\fiftyfifty$ beam splitter with phase $\varphi_\text{ent}=\pi/2$. The two output modes are entangled and measured by homodyne detection. The detected quadratures are determined by the phase $\varphi_A$ and $\varphi_B$ of the local oscillators.}
  \label{fig:sclass}
\end{figure}

CV entanglement was first observed by Ou et al.~\cite{Ou1992}, using type II parametric down-conversion and by Furusawa et al.~\cite{Furusawa1998}, using type I parametric down-conversion. Figure~\ref{fig:sclass} depicts the principle of quadrature entanglement generation using type I parametric down-conversion. In this case two squeezed vacuum modes are independently generated by two type I squeezed-light sources~\cite{Eberle2010, Mehmet2011}. In the figure they are represented by their Wigner functions, where $X$ denotes the amplitude quadrature and $P$ denotes the phase quadrature. After superimposing both modes with relative phase $\varphi_\text{ent} = \pi/2$ at a balanced beam splitter, the outputs are quadrature entangled. To verify the entanglement both modes are measured with homodyne detection, where the phases of the local oscillators, $\varphi_A$ and $\varphi_B$, determine the measured quadrature. Such a scheme was implemented for instance in Refs.~\cite{Furusawa1998,Bowen2003,Takei2006,Hage2011,Steinlechner2013}. In Ref.~\cite{Steinlechner2013} this scheme was used to generate \unit[9.9]{dB} entanglement at a wavelength of \unit[1064]{nm}, however, $\varphi_\text{ent}$ was not controlled, but inherently stable for up to about \unit[500]{ms}. 

\begin{figure}
  \centering
  \includegraphics[width=13cm]{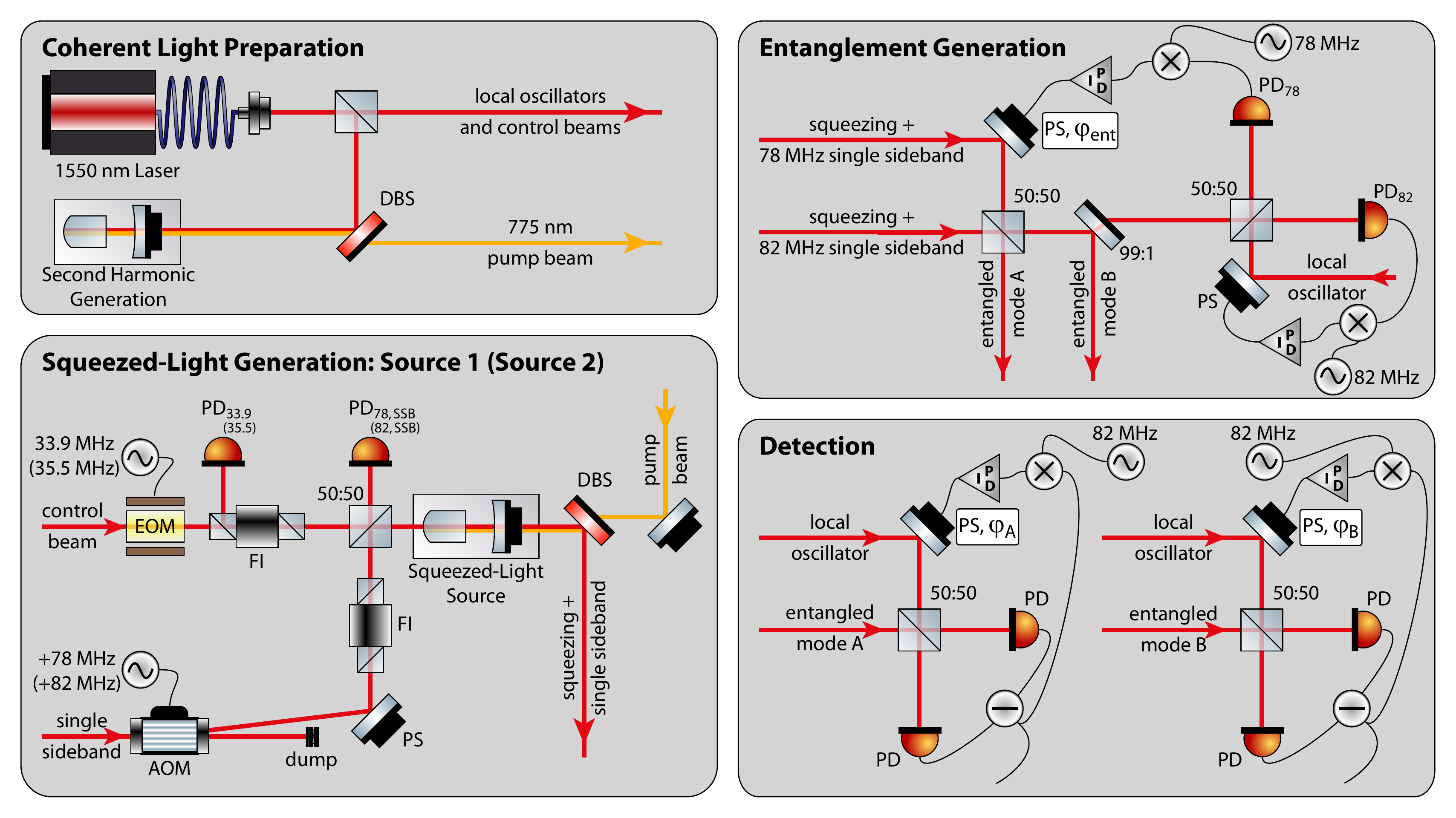}
  \caption{Schematic of the experiment. The continuous-wave fiber laser output at \unit[1550]{nm} was frequency doubled to \unit[775]{nm} which served as the pump beam for two degenerate, type I parametric squeezed-light sources. The cavity lengths of the squeezed-light sources and the phases of the pump beams were locked by means of control beams with phase modulation sidebands at \unit[33.9]{MHz} and \unit[35.5]{MHz}, respectively. A single sideband field at \unit[78]{MHz} and \unit[82]{MHz}, respectively, was locked to the control beam and served as reference for the squeezed quadratures. After superimposing the squeezed modes at a balanced beam splitter, a small fraction (\unit[1]{\%}) of one of the output modes was interfered with a local oscillator to control the phase between the two squeezed modes. The entangled modes, $A$ and $B$, were measured by balanced homodyne detection. DBS: Dichroic Beam Splitter, PD: Photo Diode, FI: Faraday Isolator, EOM: Electro-Optical Modulator, AOM: Acousto-Optical Modulator, PS: Phase Shifter.}
  \label{fig:setup}
\end{figure}

In this Letter we report a control scheme for two-mode squeezed vacuum states of light at the standard telecommunication wavelength of \unit[1550]{nm}, which allows, in particular, for a stable control of the phase angles $\varphi_\text{ent}$, $\varphi_A$ and $\varphi_B$ to arbitrary values. For generating two-mode squeezed \emph{vacuum} states the control scheme involves only auxiliary beams with low power that are shot-noise limited at the measurement Fourier frequency. It also introduces only a small amount of optical loss, which is necessary for generating highly entangled states. 

\section{Experimental Setup}

A schematic of the experiment is shown in Fig.~\ref{fig:setup}. The main laser source was a fiber laser with an output power of about \unit[1]{W} at \unit[1550]{nm}. Most of the beam was frequency doubled by second harmonic generation in a potassium titanyl phosphate (PPKTP) crystal~\cite{Eberle2010,Ast2011} and served as a pump for both squeezed-light sources. 

Both squeezed-light sources were made of a PPKTP crystal of about \unit[10]{mm} length. One end face of the crystal was curved with a radius of curvature of \unit[12]{mm} and high-reflective coated, the other was plane and anti-reflective. The cavity was formed by the curved end face of the crystal and a curved coupling mirror with a radius of curvature of \unit[25]{mm}. The coupling mirror had a reflectivity of \unit[90]{\%} for \unit[1550]{nm} and \unit[20]{\%} for \unit[775]{nm}. The temperature was controlled to the phase matching temperature by a peltier element. The length of the cavity could be controlled by a piezo-electric transducer (PZT) attached to the coupling mirror. Both the cavity length and the pump phase were controlled by using a weak control beam of about \unit[1.6]{mW} ($\unit[800]{\mu W}$ after the \fiftyfifty beam splitter) which was shot-noise limited at about \unit[5]{MHz}. Phase modulation sidebands were imprinted to the control beam by an electro-optical modulator at \unit[33.9]{MHz} for the first and \unit[35.5]{MHz} for the second squeezed-light source. The beam reflected at the cavity was split from the input beam by a Faraday isolator (FI) and detected by a resonant photo detector. The photo current was demodulated at the sideband frequency into its I and Q components which served as error signals for the cavity length~\cite{Black2001} and pump phase~\cite{Hage2007}. All optical phases in the experiment were actuated by a PZT attached to a mirror. The phase of the pump was controlled to yield deamplification of the control beam. Hence, the amplitude quadrature was squeezed. 

About $\unit[30]{\mu W}$ of the main laser beam were frequency shifted by \unit[78]{MHz} for the first and by \unit[82]{MHz} for the second squeezed-light source by means of an acousto-optical modulator (AOM). After passing a FI which prevented parasitic cavities, the beam was superimposed with the control beam at a \fiftyfifty beam splitter forming a single sideband (SSB) at the respective frequency. One output port of the beam splitter was detected by a photo detector and demodulated at the SSB frequency to generate an error signal for the phase lock of the SSB to the control beam. Hence, the SSB became a phase reference for the squeezed quadrature angle. The squeezed light leaving the cavity through the coupling mirror was split from the pump beam by a dichroic beam splitter.

While for experiments using state of the art \unit[1064]{nm} lasers the SSB can be superimposed with the squeezed field at the dichroic beam splitter behind the squeezed-light source~\cite{DiGuglielmo2011}, this procedure has several drawbacks when using \unit[1550]{nm} fiber lasers. Firstly, a phase-locked loop of a frequency shifted auxiliary laser to the main laser is difficult to achieve because of the large phase noise of fiber lasers. Secondly, when using AOMs, the amplitude modulation introduced by these devices typically has about the same size as the beat-note of the control beam transmitted through the high reflective dichroic beam spitter (\unit[99.95]{\%}) and the frequency shifted beam. Although the amplitude modulation is only visible in the amplitude quadrature, the phase quadrature cannot be used for error signal generation because of the large phase noise which reduces the signal-to-noise ratio of the already small error signal. Therefore we preferred the aforementioned superposition of the SSB with the control beam in front of the squeezed-light source which yielded a large signal-to-noise ratio for the error signal of the phase-locked loop.

Both amplitude squeezed beams were superimposed at a balanced beam splitter whose output modes are denoted as mode $A$ and mode $B$. To achieve two-mode squeezing according to Fig.~\ref{fig:sclass}, the phase $\varphi_\text{ent}$ between the squeezed beams had to be controlled to $\pi/2$. Therefore we tapped off a fraction of \unit[1]{\%} in one of the output beams of the beam splitter. The tap-off was superimposed at a balanced beam splitter with a local oscillator with a power of about \unit[5]{mW}. The output beams were detected by two resonant photo detectors, $\text{PD}_{78}$ and $\text{PD}_{82}$, and demodulated at \unit[78]{MHz} and \unit[82]{MHz}, respectively. The error signal generated by the \unit[82]{MHz} demodulation was used to control the phase of the local oscillator, while the other was used to control $\varphi_\text{ent}$. By changing the phase of the electronic local oscillator used for the demodulation, $\varphi_\text{ent}$ could be controlled to an arbitrary angle. Hence, besides the $\pi/2$ phase shift used in this experiment, $\varphi_\text{ent}$ can be locked to values more suitable in other experiments, e.g.\ for quantum dense metrology~\cite{Steinlechner2012}. Generating the error signal for $\varphi_\text{ent}$ in such a two-fold manner, increases the signal-to-noise ratio dramatically compared to the detection of the SSB-SSB beat or the beat of one of the single sidebands with the control beams. For the latter also a fraction of mode $A$  would be needed~\cite{DiGuglielmo2011}, increasing the total optical loss.

Both entangled modes were detected by balanced homodyne detection. The homodyne detector's local oscillators with a power of \unit[10]{mW} each, were spatially filtered by a polarization maintaining fiber and superimposed with mode $A$ and mode $B$, respectively, with a visibility of about \unit[99.5]{\%}. The outputs of the balanced beam splitter were detected by a pair of high quantum efficiency photo diodes. The photo current difference was demodulated at \unit[82]{MHz} to generate an error signal for the local oscillator's phase $\varphi_A$ and $\varphi_B$, respectively. Here, the measured quadrature is defined by the phase of the electronic local oscillator used for the demodulation and hence, could be controlled to arbitrary values. An amplified output of the homodyne detector electronics was anti-alias filtered and sampled by a data acquisition card with two synchronized channels with a sampling frequency of \unit[256]{MHz}. The samples were demodulated digitally at \unit[8]{MHz} and lowpass filtered at \unit[200]{kHz}. The electronic dark noise variance of the homodyne detector electronics was about \unit[20]{dB} below the vacuum noise variances, measured with blocked signal ports.

\section{Results}
\begin{figure}[ht]
  \center
  \subfloat[Duan Criterion (critical value $4$).]{\includegraphics[width=8.5cm]{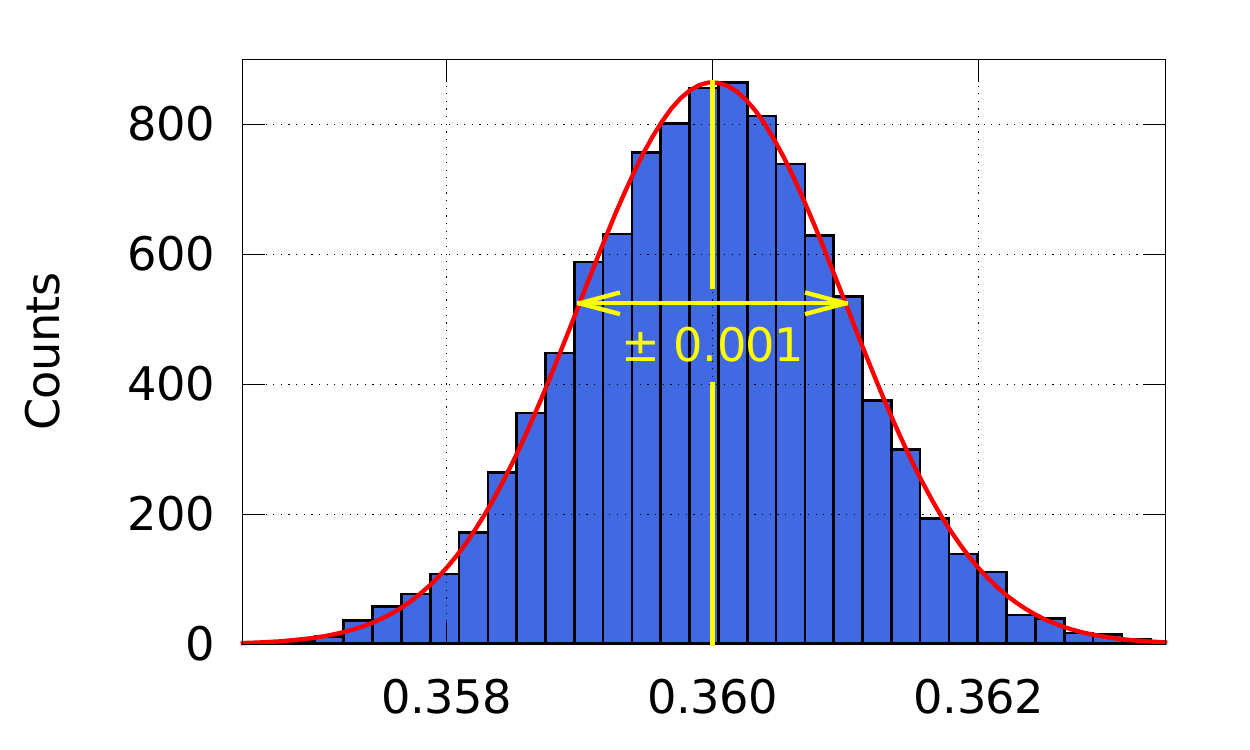}}\\
  \subfloat[EPR-Reid Criterion (critical value $1$)]{\includegraphics[width=8.5cm]{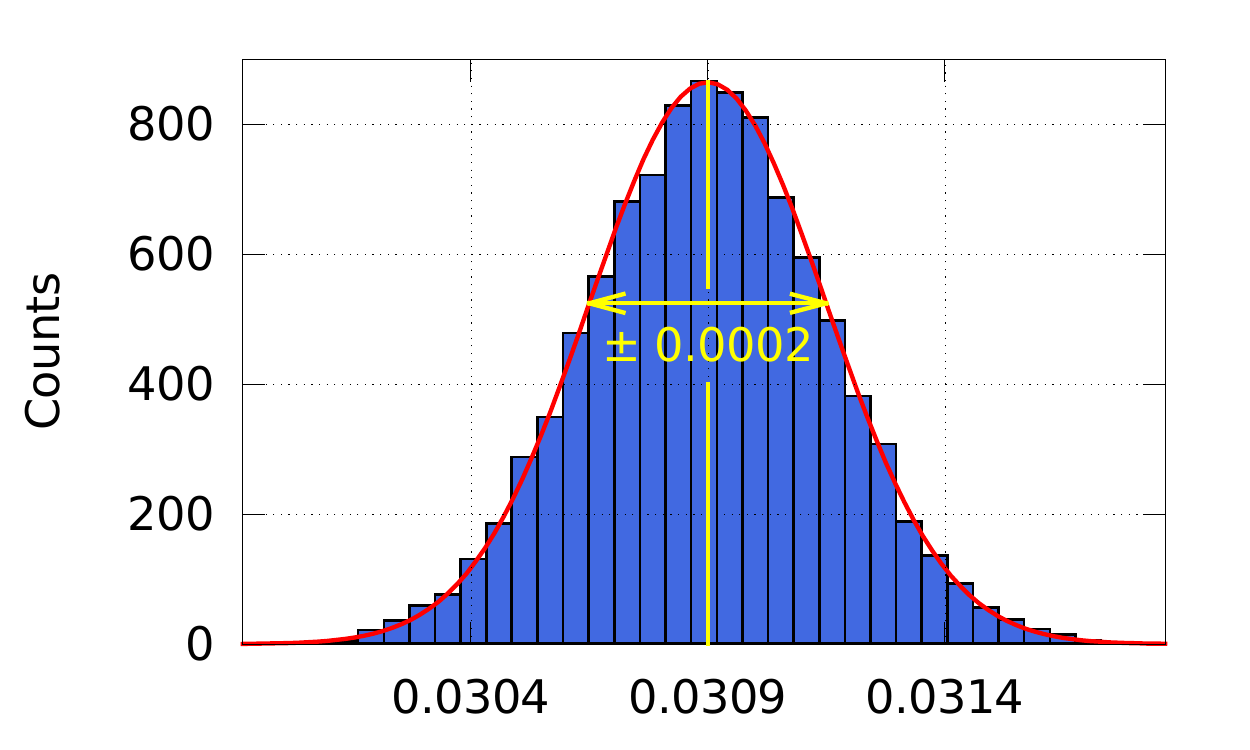}}
  \caption{Experimental Results. Histograms of the Duan inseparability and the EPR-Reid criterion with critical values of $4$ and $1$, respectively. Both histograms were computed by bootstrapping $10^6$ data points into $10^4$ chunks of $2\times10^5$ length.}
  \label{fig:results}
\end{figure}
The measured two-mode squeezed vacuum state was generated with a pump power of about \unit[0.2]{W} for each squeezed-light source. $\varphi_\text{ent}$ was controlled to $\pi/2$. The vacuum noise reference was measured by blocking the signal ports of the homodyne detectors. By controlling $\varphi_A$ and $\varphi_B$ to the amplitude or phase quadrature we made a partial tomographic measurement~\cite{DiGuglielmo2007}. For each quadrature setting we recorded $10^6$ data points, from which we reconstructed $12$ out of $16$ entries of the covariance matrix
\begin{equation}
    \gamma = \begin{pmatrix}
        21.813  &  (\approx 0)    & -21.725 & -0.010 \\
        (\approx 0)     &  25.750 & -0.140  & 26.120 \\
        -21.725 &  -0.140 & 21.801  & (\approx 0)    \\
        -0.010  & 26.120  & (\approx 0)     & 26.685 \\
    \end{pmatrix}\ .
\end{equation}
Here, the values given in brackets could not directly be measured as they correspond to non-commuting operators. In principle, these entries of the covariance matrix can be calculated from additional measurements at a linear combination of the amplitude and phase quadrature. Such measurements were first demonstrated in~\cite{DiGuglielmo2007}. Since $\varphi_{\text{ent}}$ was precisely controlled to $\pi/2$, as well as the phases of the homodyne detectors' local oscillators were precisely controlled to the amplitude and phase quadratures, however, the covariances, which were not determined, should be close to $0$~\cite{Steinlechner2013}.

The entanglement was verified by the Duan inseparability criterion for symmetric states~\cite{Duan2000}
\begin{equation}
    \Var(\op{X}_A + \op{X}_B) + \Var(\op{P}_A - \op{P}_B) < 4\ ,
\end{equation}
and the Einstein-Podolsky-Rosen (EPR) criterion by Reid~\cite{Reid1989}
\begin{equation}
  \min_g \Var(\op{X}_{\{A,B\}} - g\op{X}_{\{B,A\}})\ \times \min_h \Var(\op{P}_{\{A,B\}} - h\op{P}_{\{B,A\}}) < 1\ ,
\end{equation}
where $\Var$ denotes the variance. The results are shown in Fig.~\ref{fig:results}. The histograms were calculated by bootstrapping the $10^6$ data points into $10^4$ chunks of $2 \times 10^5$ length. A Gaussian function was fitted to the histogram yielding $0.360 \pm 0.001$ for the Duan criterion and $0.0309 \pm 0.0002$ for the EPR-Reid criterion for the $A$ to $B$ direction. For the other direction similar results were obtained. In Ref.~\cite{Steinlechner2013} $0.41$ for the Duan criterion and $0.04$ for the EPR criterion were measured at \unit[1064]{nm}, already outperforming all previous experiments on continuous variable entanglement.

While the measurements shown here were performed at \unit[8]{MHz} with a bandwidth of \unit[200]{kHz}, the spectrum of the entanglement is identical to the spectrum of the squeezed states which were used to generate the entanglement. The spectrum of our identically built squeezed-light sources can be found in Fig.~$4$ of Ref.~\cite{Mehmet2011}.

\begin{figure}[ht]
    \center
    \includegraphics[width=8.5cm]{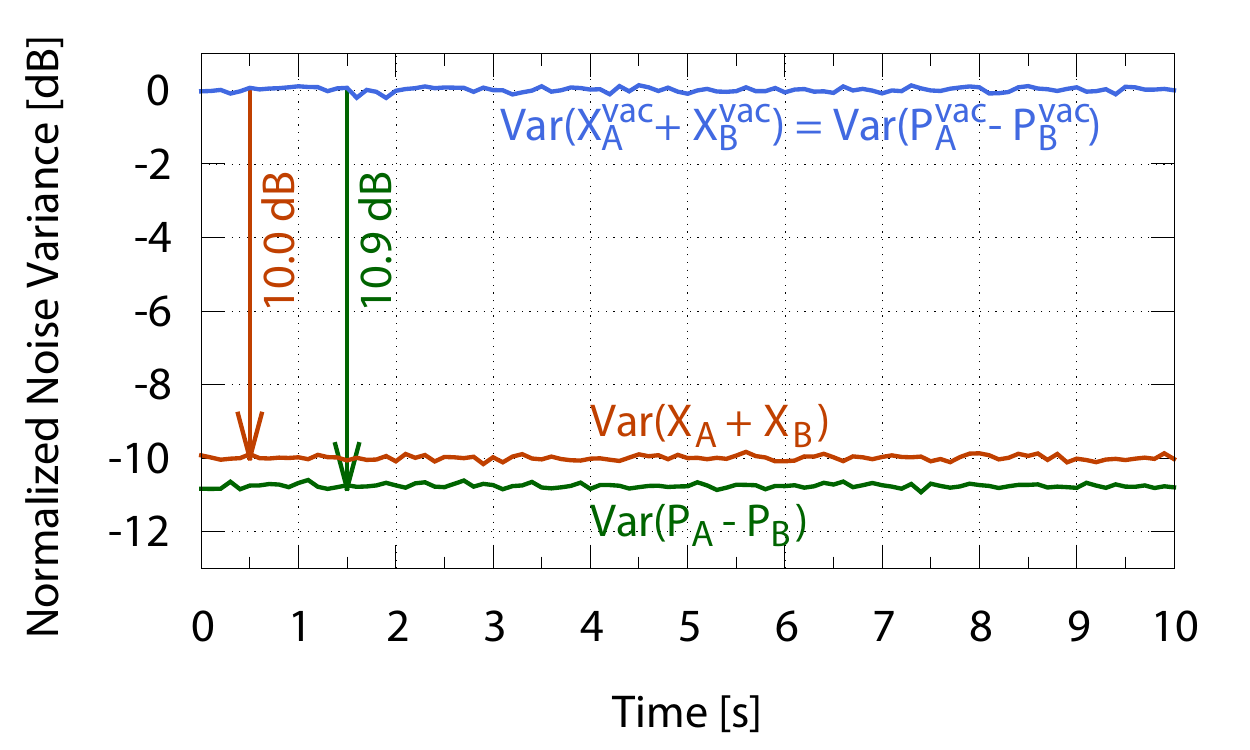}
    \caption{Stability of our entanglement source. The noise variances $\Var(\op{X}_A+\op{X}_B)$ and $\Var(\op{P}_A-\op{P}_B)$ normalized to the variance of the sum or difference of the quadratures for a vacuum state are plotted versus time. The noise variances are stable over the shown measurement time.}
    \label{fig:variances vs times}
\end{figure}
To demonstrate the stability of our active control loops, Fig.~\ref{fig:variances vs times} shows the variance of the sum of the amplitude quadrature operators, $\Var(\op{X}_A + \op{X}_B)$, and the variance of the difference of the phase quadrature operators, $\Var(\op{P}_A - \op{P}_B)$, versus time. Both variances were normalized to a joint measurement of vacuum states at the homodyne detectors, $\Var(\op{X}_A^\text{vac} + \op{X}_B^\text{vac})$ and $\Var(\op{P}_A^\text{vac} - \op{P}_B^\text{vac})$, respectively. Over the measurement time of \unit[10]{s} the noise variances were stable at about \unit[10.0]{dB} and at about \unit[10.9]{dB} for the amplitude and phase quadrature, respectively. Without our active control loops the noise suppression would reach the same values, however, only stable over short time scales. For instance, in Ref.~\cite{Steinlechner2013}, where $\varphi_\text{ent}$ was not locked, the measurement time was only $\unit[200]{\mu s}$. The stability of our phase lock was not limited to the \unit[10]{s} being presented in the figure. Indeed, we observed the stable production of our entangled states for more than \unit[15]{min}. In principle, our active control loops allow an extension of the measurement time to arbitrary duration if the dynamic ranges of the used piezo actuators are large enough to compensate for thermal drifts.

The optical loss of our squeezed-light sources was slightly asymmetric with an outcoupling efficiency of about \unit[96]{\%} for the first and about \unit[97.5]{\%} for the second source. The fringe visibility at the entangling beam splitter was about \unit[99.5]{\%}. Taking into account the \unit[1]{\%} loss introduced by the tap-off in one arm for the phase lock at the entangling beam splitter and together with a fringe visibility of about \unit[99.5]{\%} at the homodyne detectors' beam splitters, the quantum efficiency of about \unit[99]{\%} and propagation losses of about \unit[1]{\%}, the observed values for the Duan and EPR-Reid criterion are reproduced quite well. We observed no evidence for phase noise, showing the good performance of the implemented control scheme.

\section{Conclusion}

In conclusion we have demonstrated a phase control scheme for two-mode squeezed vacuum states at the telecommunication wavelength of \unit[1550]{nm}. Using this scheme we generated states which showed a Duan inseparability value of \unit[10.45]{dB} and an EPR-Reid value of $0.0309 < 1$. The demonstrated control scheme allowed for arbitrary phase angles between the squeezed modes and for arbitrary homodyne angles, while introducing only \unit[1]{\%} optical loss in one arm. No evidence for phase noise introduced by the locks was found. The observed states are highly suitable for demanding experiments like CV superactivation and a CV quantum cryptography proof-of-principle experiment showing security against most general coherent attacks including finite size effects. Indeed a positive key rate for more than $4 \times 10^7$ samples (after sifting) is feasible with the present state~\cite{Furrer2012}. For very large numbers of samples the key rate would reach about \unit[0.8]{bits/sample}. The wavelength of \unit[1550]{nm} makes the states compatible with existing telecommunication fiber networks. The good mode shape, as shown by the high visibility of \unit[99.5]{\%} at both homodyne detectors, allows high coupling efficiencies to optical fibers as demonstrated in Ref.~\cite{Mehmet2010}.

\section*{Acknowledgements}
This research was supported by the FP\,7 project Q-ESSENCE (Grant agreement number 248095) and the Centre for Quantum Engineering and Space-Time Research (QUEST). TE and VH thank the IMPRS on Gravitational Wave Astronomy for support. VH thanks HALOSTAR for financial support. We thank Torsten Franz for calculating the achievable secure key rates for our state, and Sebastian Steinlechner and Aiko Samblowski for helpful discussions.

\end{document}